\documentclass[a4paper, 10pt, conference]{ieeeconf}
\IEEEoverridecommandlockouts
\usepackage{bbm}
\usepackage{amsfonts}
\usepackage{amsmath}
\newtheorem{thm}{Theorem}[section]
\newtheorem{lem}[thm]{Lemma}

\begin{document}

\title{Control of decoherence in open quantum systems using feedback}

\author{Narayan Ganesan and Tzyh-Jong Tarn\thanks{Narayan Ganesan, {\tt\small ng@ese.wustl.edu}
and T. J. Tarn, {\tt\small tarn@wuauto.wustl.edu} are with the Department of Electrical and Systems Engineering,
Washington University in St. Louis, MO-63130 }
}


\maketitle

\begin{abstract}

Quantum feedback is assuming increasingly important role in quantum control and quantum information processing.
In this work we analyze the application of such feedback techniques in eliminating decoherence in open quantum
systems. In order to apply such system theoretic methods we first analyze the invariance properties of quadratic
forms which corresponds to expected value of a measurement and present conditions for decouplability of
measurement outputs of such time-varying open quantum systems from environmental effects.

\end{abstract}


\section{Introduction}

Decoherence is the process by which quantum systems lose their coherence information by coupling to the
environment. The quantum system entangles to the states of the environment and the system density matrix can be
diagonalized in a preferred basis states for the environment, dictated by the model of interaction
hamiltonian\cite{joos}\cite{zurek}. Decoherence is now the biggest stumbling block towards exploitation of
quantum speedup\cite{chuang} using finite quantum systems in information processing. Many authors have addressed
the control and suppression of decoherence in {\it open-quantum systems} by employing a variety of open loop and
feedback strategies.  Effect of decoherence suppression under arbitrarily fast open loop control was studied by
Viola et al \cite{viola1}\cite{viola2}. Another method along similar lines for control of decoherence by
open-loop multipulses was studied by Uchiyama et. al.\cite{uchiyama}. A very illustrating example of decoherence
of single qubit system used in quantum information processing and its effective control using pulse method was
worked out by Protopopescu et al\cite{proto}. Shor\cite{shor} and Calderbank\cite{calderbank} also came up with
interesting error-correction schemes for detecting and reducing effects of decoherence on finite quantum
registers. Recently many authors have also studied the application of feedback methods in control of
decoherence\cite{doherty},\cite{horoshko}. Technological advances enabling manipulation, control of quantum
systems and recent advances in quantum measurements using weak coupling, non-demolition
principles\cite{barginsky} etc, has opened up avenues for employing feedback based control strategies for
quantum systems \cite{wallentowitz},\cite{jacobs},\cite{horoshko}.

In this work we analyze the effectiveness of feedback method in eliminating decoherence. A wave function
approach as opposed to density matrices for the schr\"{o}dinger equation is adopted which represents the system
in an input-affine form and greatly enables one to exploit methodologies from systems theory. We first analyze
what it means for a complex scalar function to be invariant of certain parameters. The generality of the
treatment adopted here makes all types of quantum systems amenable to the results. It is also shown here that
analysis of invariance of quadratic forms also lead to Decoherence Free Subspaces (DFS) for the open quantum
systems but from a different and general perspective. DFS was first shown to exist by Lidar et al\cite{lidar} by
analysis of Markovian master equation for open quantum systems that naturally gives rise to subspaces that are
immune to the effects of decoherence namely dissipation and loss of coherence.

\section{Mathematical Preliminaries}
We explore the conditions for a scalar function represented by a quadratic form of a time varying quantum
control system to be invariant of perturbation or interaction hamiltonian when coupled to a quantum environment.

Let
\begin{eqnarray*}
&\frac{\partial\xi(t,x)}{\partial t}=&[H_0\otimes \mathcal{I}_e(t,x)+\mathcal{I}_e \otimes H_e(t,x)+H_{SB}(t,x)
\\&& +\sum_{i=1}^{r}u_i(t)H_i\otimes \mathcal{I}_e(t,x)]\xi(t,x) \label{opqusys}
\end{eqnarray*}

be the governing Schrodinger equation for a quantum system
interacting with the environment.\\
$\mathcal{H}_s$ be the system's Hilbert space.\\
$\mathcal{H}_e$ be the environment's Hilbert space.\\
$\mathcal{H}_s$ could be finite or infinite dimensional and
$\mathcal{H}_e$ is generally infinite dimensional.\\
$\xi(t,x)$ be the wave function of the system and environment.\\
$H_0$ and $H_e$ are respectively the drift Hamiltonian of the system and environment while $H_i$'s are the
control Hamiltonian of the system. $H_{SB}$ governs the interaction between the system and the environment. The
above Hamiltonian are assumed to be time varying and dependent on the spatial variable. Consider a scalar
function (typically the expected value of an observable) of the form,
\begin{equation}
y(t,\xi)=\langle \xi(t,x) | C(t,x) |\xi(t,x) \rangle \label{opeq}
\end{equation}
where again $C(t,x)$ is assumed to be time-varying operator acting on system Hilbert space. The above is the
general form of a time dependent quantum system and we wish to study the invariance properties of the function
$y(t,\xi)$ with respect to the system dynamics.

Let $y(t,\xi)=f(t,x,u_1,\cdots,u_r,H_{SB})$ be a complex scalar map of the system as a function of the control
functions and interaction Hamiltonian over a time interval $t_0\leq t\leq t_1$. The function is said to be
invariant of the interaction Hamiltonian if
\begin{equation}
f(t,x,u_1,\cdots,u_r,H_{SB}) = f(t,x,u_1,\cdots,u_r,0) \label{cond9}
\end{equation}
 for all admissible control functions $u_1,\cdots,u_r$ and a given interaction Hamiltonian $H_{SB}$.

Let $\mathcal{M}$ be the manifold contained in the Hilbert space $\mathcal{H}_s\otimes\mathcal{H}_e$ on which
the dynamics of the system is described. It could be a finite or infinite dimensional submanifold of
$\mathcal{S}_H$, the unit sphere on the collective Hilbert space. The quantum system is assumed to be governed
by time varying Hamiltonian and it is known that the system evolves on an analytic manifold
$\mathcal{D}_\omega$, which is dense in $\mathcal{M}$ and a submanifold of the unit sphere $\mathcal{S}_H$ by
Nelson's theorem\cite{htc}. Recent analysis of controllability criteria and reachability properties of states as
studied by Schirmer et.al \cite{schirmer1}\cite{schirmer2} provides insight into behavior of quantum control
systems on finite dimensional manifolds in $\mathbb{R}^n$. The controllability under various realistic
potentials was also studied by Dong et.al \cite{dong1}\cite{dong2}. However the analysis of time-varying systems
carried out here assumes in general that the component Hamiltonian operators carry explicit time dependence
which is not under the control of an external agent. And we do so by introducing a time invariant system in the
augmented state space domain $\mathcal{M}'=\mathcal{M}\oplus \mathbb{R}$. A similar scheme was also used by Lan
et.al \cite{lan} to study controllability properties of such time-varying quantum systems.

Let $x_1=t$, the new equation governing the evolution of the system can be written as,
\begin{align}
\frac{\partial}{\partial t} \left(\begin{array}{c}x_1 \\
\xi(t,x)\end{array}\right) &= \left( \begin{array}{c} 1 \\
(H_0(x_1,x)+H_e(x_1,x))\xi(t,x)
\end{array} \right) \nonumber \\
&+ \left( \begin{array}{c} 0 \\ u_i H_i(x_1,x)\xi(t,x) \end{array} \right) \label{augsys} \\
&+ \left( \begin{array}{c} 0 \\ H_{SB}(x_1,x)\xi(t,x) \end{array} \right) \nonumber
\end{align}
with,
\begin{equation}
y(t,\xi)=\langle \xi(t,x) | C(t,x) |\xi(t,x) \rangle \label{ndo}
\end{equation}

The vector fields $K_0=\left(\begin{array}{c}1\\(H_0+H_e)\xi(x,t)\end{array}\right),\\
K_i=\left(\begin{array}{c}0\\ H_i\xi(x,t)\end{array}\right)$ and
$K_I=\left(\begin{array}{c}0\\H_{SB}\xi(x,t)\end{array}\right)$ corresponding to drift, control and interaction
can be identified to contribute to the dynamical evolution.
\begin{lem}
\label{opinvlem} Consider the quantum control system (\ref{augsys}) and suppose that the corresponding output
given by equation (\ref{ndo}) is invariant under given $H_{SB}$. Then for all integers $p\geq0$ and any choice
of vector fields $X_1,\cdots,X_p$ in the set $\{K_0,K_1,\cdots,K_r\}$ we have
\begin{equation}
L_{K_I} L_{X_1} \cdots L_{X_p}y(t,\xi) = 0; \mbox{for all }t,\xi \label{cond6}
\end{equation}
\end{lem}
Before proving the above Lemma it is useful to consider a simple extension. Consider for a fixed number vector
fields $\{ X_1, \cdots, X_p \}$, with $p$ fixed and from the previous condition,
\begin{equation}
\begin{array}{cc}
L_{K_I} L_{X_1} \cdots L_{X_{p-1}}L_{K_0} y(t,\xi) = 0 & (X_p=K_0)\\
L_{K_I} L_{X_1} \cdots L_{X_{p-1}}L_{K_1} y(t,\xi) = 0 & (X_p=K_1) \\
\vdots\\
L_{K_I} L_{X_1} \cdots L_{X_{p-1}}L_{K_r} y(t,\xi) = 0 & (X_p=K_r)
\end{array}
\end{equation}
Combining the above conditions we get
\[
L_{K_I} L_{X_1} \cdots L_{X_{p-1}}L_{\tau} y(t,\xi) = 0
\] where $\tau \in \textrm{span}\{K_0,K_1,\cdots,K_r\}$. By finite mathematical
induction over all the variables we can replace the vector fields $X_1,\cdots,X_p$ with vector fields
$Z_1,\cdots,Z_p$ in $\textrm{span}\{K_0,K_1,\cdots,K_r\}$. Hence one can show that the previous condition is
equivalent to the requirement that
\begin{equation}
L_{K_I}L_{Z_1} \cdots L_{Z_p}y(t,\xi) = 0; \mbox{\it  for all $t,\xi$} \label{cond3}
\end{equation}
for all $p\geq0$ and any choice of vector fields of the form
\begin{eqnarray}
Z_i = K_0 + \sum_{j=1}^r u_j^iK_j; & u_j^i \in U \label{zform}
\end{eqnarray}
where $U$, stand for the set of admissible control functions. \\
{\it Proof }Now let $y$ be invariant under $H_{SB}$. Then for small $t_1,\cdots,t_k$ by equation (\ref{cond9})
\begin{equation}
y(t,Z_{k}^{t_{k}}\circ Z_{k-1}^{t_{k-1}}\circ \cdots \circ Z_{1}^{t_{1}}(\xi)) = y(t,\tilde{Z}_{k}^{t_{k}}\circ
\tilde{Z}_{k-1}^{t_{k-1}}\circ \cdots \circ \tilde{Z}_{1}^{t_{1}}(\xi)) \label{cond5} \\
\end{equation}
where $Z_1,\cdots,Z_k$ are of the form (\ref{zform}) and $\tilde{Z}_1,\cdots,\tilde{Z}_k$ are given by,
\begin{equation}
\tilde{Z}_k=Z_k + K_I \label{ztildeform}
\end{equation}
and $Z_{k}^{t_{k}}$, the one parameter group of flow of the vector field $Z_k$. The left hand side of equation
(\ref{cond5}) is the output for $H_{SB}=0$ while the right hand side is for an arbitrary $H_{SB}$.
Differentiating both sides of (\ref{cond5}) with respect to $t_k,t_{k-1},\cdots,t_1$ at respectively
$t_k=0,\cdots t_1=0$ yields,
\begin{equation}
L_{Z_1}L_{Z_2}\cdots L_{Z_k}y(t,\xi) = L_{\tilde{Z}_1}L_{\tilde{Z}_2}\cdots L_{\tilde{Z}_k}y(t,\xi)
\label{cond7}
\end{equation}
for all $k\geq0$. Now for $k=1$ the above equation yields,
\[
L_{Z_1} y(t,\xi)=L_{\tilde{Z}_1} y(t,\xi)
\]
Since $L_{\tilde{Z}_1}y=L_{Z_1}y + L_{K_I}y$, and using the above equation we can conclude $L_{K_I} y=0$, which
is same as the equation (\ref{cond3}) for $p=0$. Again in general by induction we obtain,
\begin{eqnarray*}
&L_{Z_1}L_{Z_2}\cdots L_{Z_k}y(t,\xi) & =
L_{\tilde{Z}_1}L_{\tilde{Z}_2}\cdots L_{\tilde{Z}_k}y(t,\xi) \\
& & = L_{\tilde{Z}_1}L_{Z_2}\cdots L_{Z_k}y(t,\xi)
\end{eqnarray*}
and using equation(\ref{ztildeform}) this yields,
\[
L_{K_I}L_{Z_1} \cdots L_{Z_k}y(t,\xi) = 0; \mbox{for all x}
\]
for all $Z_i$ of the form (\ref{zform}). The sufficient condition for output invariance however requires a
stronger condition of analyticity of the system.
 Lemma {\ref{opinvlem}} implies that the necessary conditions for output
invariance are,
\begin{align}
L_{K_I} y(t,\xi) &= 0 \nonumber{} \\
L_{K_I} L_{K_{i_0}} \cdots L_{K_{i_n}} y(t,\xi) &= 0 \label{cond7}
\end{align}
for $0\leq i_0,\cdots, i_n \leq r$ and $n\geq 0$, where $K_0,\cdots, K_r$ are the vector fields of the augmented
system and $K_I$, the interaction vector field. The previous condition can also be restated thus,
\begin{align}
L_{K_I} y(t,\xi) &= 0 \nonumber{} \\
L_{K_I} L_{\tau_{i_0}} \cdots L_{\tau_{i_n}} y(t,\xi) &= 0
\end{align}
where $\tau_{i_0}, \cdots ,\tau_{i_n} \in $ span$\{K_0,\cdots,K_r\}$. The above restatement might be helpful in
simplifying calculations for Lie derivatives.
\begin{lem}
Suppose the system (\ref{augsys}) is analytic, then $y$ is invariant under given $H_{SB}$ if and only if
(\ref{cond6}) is satisfied.
\end{lem}
{\it Proof } Consider a sequence of arbitrary control functions in $U$. Let
\begin{align*}
u(t)&=(u_1^1,\cdots,u_r^1), t\in[t_0,t_1), t_0=0;\\
&=(u_1^2,\cdots,u_r^2), t\in[t_1,t_1+t_2) \\
\vdots\\
&=(u_1^p,\cdots,u_r^p), t\in[t_1+\cdots t_{p-1},t_1+\cdots t_p) \\
\end{align*}
and two time instances $s,t$ satisfying $0\leq s\leq t\leq t_1+\cdots t_p$. We can then write,
\begin{align*}
&s=t_1+\cdots + t_{k-1}+(t_k-\tau_k)\\
&t=t_1+\cdots + t_{k-1}+t_k+\cdots + t_{l-1}+\tau_l
\end{align*}
for some index variables $k,l$ such that $0\leq k \leq l \leq p$ and some $\tau_k, \tau_l$ such that $0 \leq
\tau_k < t_k$ and $0 \leq \tau_l < t_l$. Let $\psi(t)=\left(\begin{array}{c}x_1\\ \xi(t)\end{array}\right)$ be
the state of the system in the augmented manifold and let $\gamma_0(t,s,\psi(s))$ and $\gamma_I(t,s,\psi(s))$ be
the state map of the quantum control system in the absence and presence of $K_I$ respectively, where $\psi(s)$
is the initial state at time $s$. Define a smooth function on the augmented manifold
$f(\psi)=y(\gamma_0(t,s,\psi))$. Making use of following relation,
\begin{align}
&y(t,u_1,u_2,\cdots,u_r,H_{SB})-y(t,u_1,u_2,\cdots,u_r,0) = \nonumber \\
&\int_0^t f(\psi(s))_*.K_I(\psi(s))|_{\psi(s)=\gamma_I(s,0,\psi(0))} ds \label{int}
\end{align}
Without loss of generality, considering a piecewise constant control set the term inside the integral can be
written as,
\begin{align}
&f(\psi(s))=y(\gamma_0(t,s,\psi(s))) \nonumber \\
&=y(Z_l^{\tau_l}\circ Z_{l-1}^{t_{l-1}}\circ\cdots\circ Z_{k+1}^{t_{k+1}} Z_k^{\tau_k}(\psi(s)))
\end{align}
where $Z_i$'s are of the form (\ref{zform}). Since the system was assumed to be analytic we can write,
\begin{align}
&y(Z_l^{\tau_l}\circ Z_{l-1}^{t_{l-1}}\circ\cdots\circ Z_{k+1}^{t_{k+1}} Z_{k}^{\tau_{k}}(\psi(s))) \nonumber \\
&= \sum_{i=0}^\infty \frac{\tau_l^i}{i!}L_{Z_l}^i y(Z_{l-1}^{t_{l-1}}\circ\cdots\circ Z_{k}^{\tau_{k}}(\psi(s)))
\end{align}
for some small $\tau_l,t_{l-1}, \cdots \tau_k$ such that the summation converges. The remaining terms can be
expanded in the same way for any $i$,
\begin{align}
&L_{Z_l}^i y(Z_{l-1}^{t_{l-1}}\circ\cdots\circ Z_{k}^{\tau_{k}}((\psi(s))) \nonumber \\
&=\sum_{j=0}^\infty \frac{t_{l-1}^j}{j!}L_{Z_{l-1}}^j L_{Z_l}^i y(Z_{l-2}^{t_{l-2}}\circ\cdots\circ
Z_{k}^{\tau_{k}}(\psi(s))
\end{align}
Since equation (\ref{int}) is zero for any $t\geq s\geq 0$ and any given sequence of control functions it
follows that the individual terms in the summation vanish yielding condition (\ref{cond7}) and hence as a
consequence condition (\ref{cond6}) has to hold.

\section{Invariance for the quantum system}
{\it Calculation of Lie derivatives} The Lie derivatives in the above cases can be calculated for the special
case when $\tau_1,\cdots,\tau_r \in \{K_0, K_1\cdots,K_r, K_I\}$. For instructional purposes we present here two
ways for calculating Lie derivatives of the output with respect to the vector fields of augmented system ,

\begin{align}
L_{K_I}y(t) =& \left( \begin{array}{cc}\frac{\partial \langle \xi | C(x_1) |\xi \rangle}{\partial x_1} &
\frac{\partial \langle \xi | C(x_1) |\xi \rangle}{\partial \xi}
\end{array} \right).K_I \nonumber \\
&+ K_I^* .\left(\begin{array}{c}\frac{\partial \langle
\xi | C(x_1) |\xi \rangle}{\partial x_1^*} \\
\frac{\partial \langle \xi | C(x_1) |\xi \rangle}{\partial \xi^*}
\end{array} \right)
\end{align}
where $K_I^* = \left(\begin{array}{cc} 0 & \langle \xi | H_{SB}^*
\end{array} \right)$ is the co-vector field corresponding to the vector field
$K_I$, $H_{SB}$ skew hermitian, $x_1^*$, $\xi^*$ are conjugate variables and assumed to be independent of $x_1$
and $\xi$ in calculations. Therefore $\partial \langle \xi | C(x_1) |\xi \rangle/\partial x_1^* = 0$, Hence
\begin{align}
L_{K_I}y(t) &= \left( \begin{array}{cc}\langle \xi | \dot{C(t)} |\xi \rangle & \langle \xi | C(t) \end{array}
\right).\left(\begin{array}{c} 0 \\ H_{SB}|\xi \rangle
\end{array} \right) \nonumber \\
&+ \left(\begin{array}{cc} 0 & -\langle \xi | H_{SB}
\end{array} \right)\left(\begin{array}{c}0\\
C(t) |\xi \rangle \end{array} \right) \nonumber \\
 &= \langle \xi | [C,H_{SB}] | \xi \rangle \nonumber
\end{align}
Now consider,
\begin{align}
 L_{K_0}y(t) &= \left( \begin{array}{cc}\langle \xi
| \dot{C(t)} |\xi \rangle & \langle \xi | C(t) \end{array} \right).\left( \begin{array}{c} 1 \\
(H_0+H_e)|\xi \rangle
\end{array} \right) \nonumber \\ &+ \left(\begin{array}{cc} 1 & -\langle \xi |
(H_0+H_e)\end{array} \right)\left(\begin{array}{c}\frac{\partial \langle \xi | C(x_1)
|\xi \rangle}{\partial x_1^*}\\
C(t) |\xi \rangle \end{array} \right) \nonumber \\
 &= \langle \xi |\dot{C}+ [C,(H_0+H_e)] | \xi \rangle \nonumber
\end{align}
The variable $x_1$ is replaced with $t$ as it was only a dummy variable used for calculations. Another approach
follows directly from the geometrical interpretation of Lie derivatives of scalar functions,
\[
L_{K_i}y(t) = \lim_{s \rightarrow t} \frac{d}{ds}\langle \xi | C(x_1) |\xi \rangle
\]
with only the vector field $K_i$ turned on for $i=\{0,1,\cdots,r,I\}$ (i.e)
\[
\frac{\partial}{\partial t}\left(\begin{array}{c} x_1 \\ \xi(t,x)
\end{array} \right) = K_i
\]
From straight forward calculations one obtains,
\begin{align}
L_{K_0}y(t) =& \lim_{s \rightarrow t} \frac{d}{ds}\langle \xi | C(x_1) |\xi \rangle \nonumber \\
&= \langle \dot{\xi}|C(x_1)|\xi\rangle + \langle \xi|\dot{C}(t)|\xi\rangle
+ \langle \xi|C(x_1)|\dot{\xi}\rangle \nonumber \\
&=\langle \xi |\dot{C}(t)+ [C(t),(H_0+H_e)(t,x)] | \xi \rangle
\end{align}
and similarly $ L_{K_i}y(t) = \langle \xi | [C(t),H_i(t,x)] | \xi \rangle$ for $i=\{1,\cdots,r,I\}$ and
$H_I=H_{SB}$. Following the above trend for a few Lie derivatives with respect to the vector fields
$K_0,K_1,\cdots,K_r,K_I$,
\begin{align*}
&L_{K_I}y=\langle\xi|[C,H_{SB}]|\xi\rangle=0\\
&L_{K_I}L_{K_i}y=\langle\xi|[[C,H_i],H_{SB}]|\xi\rangle=0\\
\end{align*}
\begin{align*}
L_{K_I}L_{K_i}L_{K_0}y=&\langle\xi|[[\dot{C},H_i],H_{SB}]\\+&[[[C,H_0],H_i],H_{SB}]|\xi\rangle=0\\
\end{align*}
\begin{align*}
&L_{K_I}L_{K_0}L_{K_0}y=0\\
&=L_{K_I}\langle\xi|[\dot{C},H_0]+[[C,H_0],H_0] +\frac{d^2}{dt^2}C(t) \\
&+\frac{d}{dt}[C(t),H_0(t)]|\xi\rangle=0\\
&\mbox{i.e }\langle\xi|[[\dot{C},H_0],H_{SB}]+[[[C,H_0],H_0],H_{SB}]\\
&+[\ddot{C(t)},H_{SB}]+[\frac{d}{dt}[C(t),H_0(t)],H_{SB}]|\xi\rangle=0\\
&L_{K_I}L_{K_0}L_{K_i}y=L_{K_I}\langle\xi|\frac{d}{dt}[C,H_i]+[[C,H_i],H_0]|\xi\rangle=0\\
&\mbox{i.e }\langle\xi|[\frac{d}{dt}[C,H_i],H_{SB}]+[[[C,H_i],H_0],H_{SB}]|\xi\rangle=0
\end{align*}
We are now ready to state the condition for output invariance of non-demolition measurements with respect to
perturbation or interaction Hamiltonian.
\begin{thm}
Let
\begin{align*}
&\tilde{C}_1=\mbox{span}\{ad^j_{H_i}C(t)|j=0,1,\ldots;i=1,\ldots,r\}\\
&\mathcal{C}_1=\left\{ \left(ad_{H_0}+\frac{\partial}{\partial t}\right)^j\tilde{C}_1; j=0,1,\cdots \right\}\\
&\tilde{C}_2=\mbox{span}\{ad^j_{H_i}\mathcal{C}_1(t)|j=0,1,\ldots;i=1,\ldots,r\}\\
&\mathcal{C}_2=\left\{ \left(ad_{H_0}+\frac{\partial}{\partial t}\right)^j\tilde{C}_2; j=0,1,\cdots \right\}\\
&\vdots\\
&\tilde{C}_n=\mbox{span}\{ad^j_{H_i}\mathcal{C}_{n-1}(t)|j=0,1,\ldots;i=1,\ldots,r\}\\
&\mathcal{C}_n=\left\{ \left(ad_{H_0}+\frac{\partial}{\partial t}\right)^j\tilde{C}_n; j=0,1,\cdots \right\}\\
&\vdots
\end{align*}
Define a distribution of quantum operators, $
\tilde{\mathcal{C}}(t)=\Delta\{\mathcal{C}_1(t),\mathcal{C}_2(t),\cdots{},\mathcal{C}_n(t),\cdots{}\}$. The
output equation (\ref{opeq}) of the quantum system is decoupled from the environmental interactions if and only
if,
\begin{equation}
[\tilde{\mathcal{C}}(t), H_{SB}(t)]=0 \label{ic}
\end{equation}
\end{thm}
{\it Proof  } The proof follows by noting the equivalence of equation (\ref{cond7}) with the above condition.
Consider the following term $L_{K_{i_0}} \cdots L_{K_{i_k}} y(x)$ for any $k \geq 1 $, and $i_0,\cdots, i_k \in
\{0,\cdots r\}$. From the calculations above it is the expected value of an operator of Lie brackets of
$H_{i_0},H_{i_1},\cdots H_{i_r}, C $ and their time derivatives. In particular for $k=0$, and
\[
L_{K_{i_0}}y=\langle\xi|[C,H_{i_0}]+\delta(i_0)\frac{d}{dt}C|\xi\rangle
= \langle\xi|T_1|\xi\rangle \\
\]
where $\delta(i_0)$ is the delta function that takes value $1$ when $i_0=0$ and the operator $T_1$ as defined is
such that $T_1 \in \mathcal{C}_1$. Similarly for $k=1$ we have
\begin{align*}
L_{K_{i_1}}L_{K_{i_0}}y &=\langle\xi|[[C,H_{i_0}],H_{i_1}]+[\delta(i_0)\frac{d}{dt}C,H_{i_1}] \\
&+\delta(i_1)\frac{d}{dt}([C,H_{i_0}]+\delta(i_0)\frac{d}{dt}C)|\xi\rangle \\
&= \langle\xi|T_2|\xi\rangle
\end{align*}
and $T_2 \in \mathcal{C}_2$. Continuing so, in general we have $T_n \in \mathcal{C}_n$. And by using condition
(\ref{cond7}), we have $[H_{SB},T_n] = 0$ in general for decoupling. Since the condition is true for any $n \geq
0$ and any $T_n$ and since the vector space of bounded linear operators is complete we have
$[H_{SB},\sum_{i=0}^\infty \alpha_i T_i]=\sum_{i=0}^\infty \alpha_i[H_{SB},T_i]=0$ for $\alpha_i \in
\mathbb{R}$. The converse is true by noting that any operator in the distribution $\mathcal{C}$ (i.e) for any $T
\in \mathcal{C}$ can be decomposed into a sum of operators $\sum\alpha_i T_i$ for $T_i \in \mathcal{C}_i$ and
given $[H_{SB},\sum_{i=0}^\infty \alpha_i T_i]=0 \forall \alpha_i$ which is true only when $[H_{SB},T_n] =0$ for
any $n$. Hence from the previous equations $L_{K_I}L_{K_{i_n}}L_{K_{i_{n-1}}}\cdots L_{K_{i_0}}=0$ for
$i_0,\cdots, i_k \in \{0,\cdots r\}$.

\section{Examples}

Decoherence as studied by many authors\cite{omnes}\cite{zurek}\cite{joos}, entangles the states of the system
and the environment and amounts to forcible collapse of the wave function corresponding to preferred pointer
basis decided by the environment. The evolution of such a system can only be described at best at a statistical
level.

We present two qualitatively different examples to illustrate the applicability of the above formalism in
practical quantum control systems.

\subsection{Electro-optic Amplitude Modulation}
Consider a driven electromagnetic system in a single mode subject to decoherence. The control system describing
the oscillator under the semiclassical approximation is
\begin{align*}
\frac{d}{dt}\psi(t)=&(\omega a^{\dagger} a + \sum_j \omega_j c_j^{\dagger}c_j + i u(t)(a^{\dagger} -a) \\
&+ a\sum_{j}\kappa_j^*c_j + a^{\dagger}\sum_{j}\kappa_j c_j)\psi(t)
\end{align*}

where the system represented by mode $a$ is coupled to a bath of infinite number of oscillators, $c_j$ with
corresponding coupling constants $\kappa_j$ and where $\psi(t)$ is the combined wave function of the system and
bath. The control $u(t)$ is the strength of the input current and let $H_0=\omega a^\dagger a+\sum_j \omega_j
c_j^{\dagger}c_j$ and $H_1=(a^\dagger -a)$. Let the system be monitored by a non-demolition observable
\begin{align*}
C(t)=a\exp(i\omega t)+a^\dagger\exp(-i\omega t)
\end{align*}
with the corresponding output given by $y(t)=\langle\psi(t)|C(t)|\psi(t)\rangle$. Following theorem 3.1, we have
$[C(t),H_1]=\mathrm{e}^{i\omega t} + \mathrm{e}^{-i\omega t} = 2\cos(\omega t)$ with vanishing higher order
commutators. Hence $\tilde{\mathcal{C}_1}=\{c_1*C+c_2*\mathbb{I}*\cos(\omega t), \forall c_1,c_2 \in
\mathbb{R}\}$ and since $[C(t),H_0] + {\partial C}/{\partial t} = 0$ we have $\mathcal{C}_1=\tilde{C}_1$ and the
sequence converges to $\mathcal{C}_1$ which in general need not converge at all. Since the commutator of the
interaction hamiltonian $H_{SB}=a\sum_{j}\kappa_j^*c_j + a^{\dagger}\sum_{j}\kappa_j c_j$ with the elements of
the set $\mathcal{C}_1$ are not all zero, condition ($\ref{ic}$) is not fulfilled and the non-demolition
measurement is $(i)$ not invariant of the interaction hamiltonian, $(ii)$ no longer back action evading due to
the presence of the interaction. The measurement of the observable $C(t)$ would thus reveal information about
the decoherence of the system.

\subsection{Decoherence free subspaces(DFS)} The techniques developed in the previous sections can be applied to the
problem of analyzing the decoherence free subspaces(DFS) discussed in \cite{lidar}. Decoherence free subspaces
(DFS) camouflage themselves so as to be undetected by the interaction hamiltonian due to degeneracy of their
basis states with respect to $H_{SB}$.

{\em Decoherence of a collection of 2-level systems:} For a collection of 2-level systems interacting with a
bath of oscillators the corresponding hamiltonian is
\[
H=\frac{\omega_0}{2}\sum_{j=1}^{N}\sigma_3^{(j)} + \sum_k \omega_k b_k^\dagger b_k + \sum_k\sum_{j=1}^{N}
\sigma_3^{(j)}(g_{k}b_k^\dagger + g_{k}^*b_k)
\]
where the system is assumed to interact through the collective operator $\sum_j \sigma_3^{(j)}$ and $g_k$'s
describe coupling to the mode $k$. An inquiry into what information about the system is preserved in the
presence of the interaction could be answered by expressing the operator $C$ acting on the system Hilbert space
in its general form in terms of the basis projection operators,
\begin{align*}
C(t)=\sum_{i,j=0..2^N-1}c_{ij}|i\rangle\langle j|
\end{align*}
and solving for condition (\ref{ic}). For a simple N=2 system we have after straight forward calculations
\begin{align*}
\tilde{\mathcal{C}}=\mathrm{span}\{&\sum_{i,j} c_{ij}|i\rangle\langle
j|.(j^{(1)}-i^{(1)}+j^{(2)}-i^{(2)})^K ,\\
&\forall K=0,1,2...\}
\end{align*}
 where $j^{(l)}$ etc., stands for the $l^{th}$ letter (either $0$ or $1$) of the binary word $j$.
 Condition (\ref{ic}), which is $[\tilde{\mathcal{C}},H_{SB}]=0$ now translates to
\begin{align*}
\sum_{i,j} c_{ij}|i\rangle\langle j|.(j^{(1)}-i^{(1)}+j^{(2)}-i^{(2)})^K =0, \forall K=1,2,3...
\end{align*}
or nontrivially, $j^{(1)}+j^{(2)}=i^{(1)}+i^{(2)}$, or that the two words have equal number of $1's$.

The above calculations are valid for any finite $N$, a specific example for $N=3$ is $C=|000\rangle\langle
000|+|001\rangle\langle 001|+|010 \rangle\langle 100|+|011\rangle\langle 101|$. Of particular interest are terms
like $|011\rangle\langle 101|$ and $|010\rangle\langle 100|$ as the corresponding
$y(t)=\langle\psi(t)|C(t)|\psi(t)\rangle$ which is a function of the coherence between the basis states
$|011\rangle, |101\rangle$ and $|010\rangle,|100\rangle$ is predicted to be invariant under the interaction. It
is worth noting that the operator $C(t)$ acting on system Hilbert space here need not necessarily be hermitian
and only describes preserved information in a loose sense.

 {\em Decoherence in the presence of control:} In the presence of the external controls $H_i=
u_i\sigma_1^{(i)}$, the invariance condition is no longer satisfied for the operator $C$ as
$[[C,\sigma_1^{(i)}],\sigma_3^{(j)}]\neq 0$ and hence the coherence between the states is not preserved. This is
because of the transitions outside DFS caused by the control hamiltonian. The above formalism is helpful in
analyzing general class of information that would be preserved in the presence of interaction hamiltonian which
in turn would tell us about how to store information reliably in a quantum register. Though the procedure
outlined above to determine $C(t)$ could get computationally intensive even for modest systems it is
nevertheless helpful in learning about any ansatz $C(t)$.
\section{Feedback Control}
The technique of using feedback has been considered by a number of authors \cite{doherty}, \cite{wallentowitz},
\cite{horoshko} etc. Although one cannot extract information from a quantum system without disturbing it to some
extent, due to rapid advances in quantum control technology a good deal of work carried out on weak
measurements\cite{mensky}, probabilistic state estimators\cite{openqusys}, non-demolition measurements and
filters\cite{caves}\cite{barginsky}\cite{ong2} that prevent systematic back action on the system, enable us to
extract information with minimal disturbance and can now be applied to practical quantum systems. In reality a
system is coupled to a probe which in turn is immersed in the environmental bath in order to extract state
information of the system. The effects of feedback and probe coupling are currently being investigated by the
authors under this framework. In this section we analyze the effects of minimal back action feedback on the
control of decoherence problem and derive conditions for decouplability.

Consider the augmented system equation (\ref{augsys}) that describes a time dependent quantum system and a
feedback of the form $u=\alpha(\xi)+\beta(\xi).v$ in order to preserve the input-affine structure of the state
equation, where $\alpha,\beta$ are $r\times 1$ vector and $r\times r$ matrix respectively of scalar functions
depending on state $|\xi\rangle$ of the system.
\begin{align}
\frac{\partial}{\partial t} \left(\begin{array}{c}x_1 \\
\xi(t,x) \end{array}\right) &= \left( \begin{array}{c} 1 \\
(H_0+H_e+\sum \alpha_i H_i)(x_1,x)\xi(t,x)
\end{array} \right) \nonumber \\
&+ \left( \begin{array}{c} 0 \\ \sum v_i \sum \beta_{ij}H_j(x_1,x) \xi(t,x) \end{array} \right) \nonumber \\
&+ \left( \begin{array}{c} 0 \\ H_{SB}(x_1,x)\xi(t,x) \end{array} \right) \label{fbsys}
\end{align}
where again the following vector fields can be identified $\tilde{K}_0=\left(\begin{array}{c}1\\(H_0+H_e+\sum
\alpha_i H_i)\xi(x,t)\end{array}\right), \tilde{K}_i=\left(\begin{array}{c}0\\ \sum
\beta_{ij}H_j\xi(x,t)\end{array}\right)$ and $K_I=\left(\begin{array}{c}0\\H_{SB}\xi(x,t)\end{array}\right)$.

As stated above the necessary and sufficient conditions for a scalar function $y(t)$ of the system to be
invariant of the interaction vector field is,
\begin{align}
L_{K_I} y(t) &= 0 \nonumber{} \\
L_{K_I} L_{\tilde{K}_{i_0}} \cdots L_{\tilde{K}_{i_n}} y(t) &= 0 \label{cond7fb}
\end{align}
for $0\leq i_0,\cdots, i_n \leq r$ and $n\geq 0$. Translating the above conditions into operators for the above
system we obtain the following conditions. In the equations below we omit the summation symbol and following
Einstein's convention a summation has to be assumed where ever a pair of the same index appears.
\begin{align*}
&L_{K_I}y=\langle\xi|[C,H_{SB}]|\xi\rangle=0\\
&L_{K_I}L_{\tilde{K}_i}y=\langle\xi|[[C,\beta_{ij}H_j],H_{SB}]+[C,H_j]L_{K_I}\beta_{ij}|\xi\rangle=0\\
\end{align*}
\begin{align*}
L_{\tilde{K}_i}L_{\tilde{K}_0}y=&\langle\xi|[\dot{C},\beta_{il}H_l]
+[[C,H+\alpha_j H_j], \beta_{il}H_l] \\ &+[C,H_j]L_{\tilde{K}_i}\alpha_j|\xi\rangle=0\\
\end{align*}
\begin{align}
&L_{K_I}L_{\tilde{K}_i}L_{\tilde{K}_0}y \nonumber \\
=&\langle\xi|[[\dot{C},\beta_{il}H_l],H_{SB}]+[[C,H_j]L_{\tilde{K}_i}\alpha_j,H_{SB}] \nonumber \\
&+[[[C,H_0+\alpha_j H_j], \beta_{il}H_l],H_{SB}] \nonumber \\
&+[\dot{C},H_l]L_{K_I}\beta_{il}+[C,H_j]L_{K_I}L_{\tilde{K}_i}\alpha_j \nonumber \\
&+[[C,H_0],H_l]L_{K_I}\beta_{il}+[[C,H_j],H_l]L_{K_I}\alpha_j\beta_{il}|\xi\rangle \nonumber \\
=&0 \label{egeq}
\end{align}
The first two lines of RHS of the above equality is found to belong to the distribution
$[\tilde{\mathcal{C}}(t),H_{SB}]$ and the last two lines belong to $\tilde{\mathcal{C}}(t)$. The above
calculation can be extended to any number of terms to encompass the result. In general one finds that, in the
presence of feedback terms the condition for decouplability is relaxed to
\begin{equation}
[\tilde{\mathcal{C}}(t),H_{SB}]\subset\tilde{\mathcal{C}}(t) \label{condfb}
\end{equation}
In order to solve eq.(\ref{egeq}) and consequently (\ref{condfb}) for the feedback parameters, it has to be
noted that the first two lines and last two lines of eq.(\ref{egeq}) denote operators acting on different
Hilbert spaces, namely the system-environment and just the system respectively and the two terms cannot be
reconciled unless they vanish individually which leads us back to original conditions for open loop invariance.

In other words, in order for the feedback to be an effective tool in solving the decoherence problem, the
control hamiltonians $H_i$ have to act non-trivially on both the Hilbert spaces which would enable all the
operators in (\ref{egeq}) act on system-environment Hilbert space.
\section{Invariant Subspace}
As stated above the fundamental conditions for invariance were,
\begin{align*}
L_{K_I} y(t,\xi) &= 0 \\
L_{K_I} L_{K_{i_0}} \cdots L_{K_{i_n}} y(t,\xi) &= 0
\end{align*}
where $0\leq i_0,\cdots, i_n \leq r; n\geq 0$. We now explore a larger class of vector fields $K_\tau$
containing $K_I$ that also satisfy the above conditions, i.e,
\begin{align}
L_{K_\tau} y(t,\xi) &= 0 \label{invsp}\\
L_{K_\tau} L_{K_{i_0}} \cdots L_{K_{i_n}} y(t,\xi) &= 0 \nonumber
\end{align}
Set of such vector fields form a vector space or a distribution and constitute a invariant distribution in the
sense described by the following theorems.\\

{\it Definition} The vector field $K_\tau$ satisfying equations (\ref{invsp}) is said to be in the orthogonal
subspace of the observation space spanned by the co-vector fields $dy(t,\xi), dL_{K_{i_0}} \cdots L_{K_{i_n}}
y(t,\xi), \cdots$ for all $0\leq i_0,\cdots, i_n \leq r$ and $n\geq 0$. Denoted by $K_\tau \in
\mathcal{O}^\perp$

\begin{lem}
The distribution $\mathcal{O}^\perp$ is invariant with respect to the vector fields $K_0,\cdots,K_r$ under the
Lie bracket operation. (i.e) if $K_\tau \in \mathcal{O}^\perp$, then $[K_\tau,K_i] \in \mathcal{O}^\perp$ for
$i=0,\cdots, r$ \end{lem} {\it proof: } Assuming a form for the vector field $K_\tau = \left( \begin{array}{c} 0
\\ H_\tau\xi \end{array} \right) $, the Lie bracket of $[K_\tau,K_i]$ for $i=1,\cdots,r$ can be computed as
follows,
\begin{align*}
&[K_\tau,K_i] =\left[ \begin{array}{cc}
                    0 & 0 \\ \dot{H}_i |\xi\rangle & H_i
                    \end{array} \right]
                    \left(\begin{array}{c}
                    0 \\
                    H_\tau |\xi\rangle\end{array}\right) \\
&                   - \left[\begin{array}{cc}
                    0 & 0 \\
                    \dot{H}_\tau |\xi\rangle & H_\tau \end{array} \right]
                    \left(\begin{array}{c}
                    0 \\
                    H_i|\xi\rangle
                    \end{array} \right)
= \left ( \begin{array}{c}
                0 \\
                {}[H_\tau, H_i ] |\xi \rangle
                \end{array} \right )
\end{align*}

Now using Jacobi identity,
\begin{align*}
L_{[K_\tau,K_i]}y(t)
        &=\langle\xi|[C,[H_\tau,H_i]]|\xi\rangle\\
        &=-\langle\xi|[H_\tau,[H_i,C]]|\xi\rangle -\langle\xi|[H_i,[C,H_\tau]]|\xi\rangle\\
        &=-L_{K_\tau}L_{K_i}y(t)-L_{K_i}L_{K_\tau}y(t)\\
        &=0
\end{align*}
Now for $i=0$ and $K_0=\left(\begin{array}{c}1 \\
H_0|\xi\rangle \end{array} \right)$ we have,

\begin{align}
&[K_\tau,K_0]\\
  & = \left[\begin{array}{cc}
             0 & 0 \\
             0 & H_0
             \end{array} \right]
             \left(\begin{array}{c}
              0 \\
              H_\tau |\xi\rangle
              \end{array}\right)
              - \left[\begin{array}{cc}
              0 & 0 \\
              \dot{H}_\tau | \xi \rangle & H_\tau \end{array} \right]
              \left(\begin{array}{c}
              1 \\
              H_0|\xi\rangle \end{array} \right) \nonumber \\
  & = \left ( \begin{array} {c}
             0 \\
             ([H_\tau,H_0 ] -\dot{H}_\tau ) | \xi \rangle
             \end{array} \right) \nonumber \\
&L_{[K_\tau,K_0]}y(t)
   =  \langle\xi|[C,[H_\tau,H_0]] -
          [C,\dot{H}_\tau]|\xi\rangle \label{inveq}
\end{align}
We already have,
\begin{eqnarray*}
L_{K_\tau}L_{K_0}y(t,\xi)
  &=& \langle\xi|[\dot{C},H_\tau]+[[C,H_0],H_\tau]|\xi\rangle=0\\
L_{K_0}L_{K_\tau}y(t,\xi)
  &=& \langle\xi|\frac{d}{dt}[C,H_\tau]+[[C,H_\tau],H_0]|\xi\rangle=0
\end{eqnarray*}
Adding the above equations and using Jacobi Identity we conclude that $[K_\tau,K_0] \in \mathcal{O}^\perp$.

\section{Conclusion}

We analyzed the conditions for eliminating the effects of decoherence on quantum system whose coherence can be
monitored in the form of a scalar output equation. The results hold globally on the analytic manifold.

\section{Future Work}
The invariant distributions possess many desirable qualities and helps in control of decoherence. We wish to
construct an algorithm to determine the invariant distribution for a given quantum system and its interactions.
Design and study of feedback and analysis of the resulting stability for quantum control system will help us
solve the decoherence problem for practical quantum systems. The results can be extended and conditions can be
derived for different types of measurements and information extraction schemes.
\section{ACKNOWLEDGMENTS}

This research was supported in part by the U. S. Army Research Office under Grant W911NF-04-1-0386. T. J. Tarn
would also like to acknowledge partial support from the China Natural Science Foundation under Grant Number
60433050 and 60274025. The authors would also like to thank the reviewers for their invaluable comments and
suggestions.


\end{document}